# Resistive transition of hydrogen-rich superconductors

Evgeny F. Talantsev[1,2] and Karoline Stolze[3]

[1]M.N. Mikheev Institute of Metal Physics, Ural Branch, Russian Academy of Sciences, 18, S. Kovalevskoy St., Ekaterinburg, 620108, Russia

[2]NANOTECH Centre, Ural Federal University, 19 Mira St., Ekaterinburg, 620002, Russia

[3]Leibniz-Institut für Kristallzüchtung, Max-Born-Straße 2, 12489 Berlin, Germany

**Abstract**

Critical temperature, $T_c$, and transition width, $\Delta T_c$, are two primary parameters of the superconducting transition. The latter parameter reflects the superconducting state disturbance originating from the thermodynamic fluctuations, atomic disorder, applied magnetic field, the presence of secondary crystalline phases, applied pressure, etc. Recently, Hirsch and Marsiglio (2020 arXiv:2012.12796) performed an analysis of the transition width in several near-room-temperature superconductors (NRTS) and reported that the reduced transition width, $\Delta T_c/T_c$, in these materials does not follow a conventional trend of transition width broadening on applied magnetic field observed in low- and high-$T_c$ superconductors. Here we present thorough mathematical analysis of the magnetoresistive data, $R(T,B)$, for the high-entropy alloy $(ScZrNb)_{0.65}[RhPd]_{0.35}$ and hydrogen-rich superconductors of $Im$-$3m$-$H_3S$, $C2/m$-$LaH_{10}$ and $P6_3/mmc$-$CeH_9$. We found that the reduced transition width, $\Delta T_c/T_c$, in these materials does follow a conventional broadening trend on applied magnetic field.

# Resistive transition of hydrogen-rich superconductors

## I. Introduction

In 1970 Satterthwaite and Toepke [1] formulated a conceptual idea which became mainstream research in superconductivity 45 years later: "…There has been theoretical speculation [2] that metallic hydrogen might be a high-temperature superconductor, in part because of the very high Debye frequency of the proton lattice. With high concentrations of hydrogen in the metal hydrides one would expect lattice modes of high frequency and if there exists an attractive pairing interaction one might expect to find high-temperature superconductivity in these systems also." In 2004 Ashcroft [3] proposed rigorous mathematical routine for the idea.

Satterthwaite and Toepke [1] also showed that their conceptual idea can be proved in experiment. By 1970 it had been already known (mainly by extended experimental work performed at Bell Telephone Laboratories) that conventional metallic hydrides/deuterides $VH_{0.5}$ [4], $NbH_x$ (x = 0.88, 0.99), $NbD_y$ (y = 0.11, 0.13, 0.79, 0.80) [5], $ZrH_2$ [6], $TiH_2$ [6], and, remarkably, $LaH_{2.45}$ [7] and dihydride of thorium, $ThH_2$ [6], are not superconductors. Thus, Satterthwaite and Toepke [1] searched for metallic superhydrides (or, in their words, "high hydrides") and they found that thorium superhydride, $Th_4H_{15}$, and its isotopic counterpart $Th_4D_{15}$, are both superconductors with the transition temperature of $T_c$ = 8.0-8.3 K. In addition, they found that a prominent isotope effect in $Th_4H_{15}$ vs $Th_4D_{15}$ phases was not observed [1], which indicates a departure from the expected electron-phonon mediated mechanism of superconductivity [8,9]. Wang *et al* [10] reported on the discovery of a low-$T_c$ phase of $Th_4H_{15}$ which exhibits $T_c$ ~ 6 K.

It took several decades (from pivotal prediction by Satterthwaite and Toepke [1]) for Drozdov *et al* [11] to show that superhydrides can be synthesized at high pressure ($P$ > 100 GPa) and high temperature ($T$ > 1000 K) inside of a diamond anvil cell (DAC). To date more than a dozen superhydride phases, including ones exhibiting near-room-temperature superconductivity (NRTS), have been synthesized from Pr-H [12], P-H [13], Pt-H [14], Sn-H [15], Ce-H [16], Th-H [17], S-(H,D) [11,18-22], Y-H [23,24], La-(H,D) [25-29], Ba-H [30] binary systems. Recently, Semenok *et al* [31] reported on the first observation of the superconductivity in a La-Y-H ternary system.

A primary experimental technique to study NRTS is the magnetoresistance measurement, $R(T,B)$. In many reports, this measurement is accompanied or initiated by first-principle calculations [32-40]. The latter is a modern research tool which uses computing power to predict thermodynamically stable phases, density of states, phonon spectra, superconducting transition temperatures and some other superconducting parameters (like, the superconducting gap energy, the coherence length, the Ginzburg-Landau parameter, critical fields, etc.).

However, some features of the superconducting transition, and particularly the transition width, $\Delta T_c$, remains to be revealed by experimental data analysis only. In this regard, we can mention a recent report by Hirsch and Marsiglio [41] who analysed $R(T,B)$ curves in NRTS materials and found that the reduced transition width, $\frac{\Delta T_c}{T_c}$, in these compounds is independent from an applied magnetic field within the range of $0 \leq \frac{B_{appl}}{B_{c2}(0)} \leq 0.15$, where $B_{c2}(0)$ is the ground state upper critical field. This dependence is different from the one in Nb$_3$Sn [42], K$_3$C$_{60}$ [43], NbN [44], MgB$_2$ [45], YBa$_2$Cu$_3$O$_{7-\delta}$ [46], BaFe$_{2-x}$Ru$_x$As$_2$ (x = 0.71) [47], La-doped CaFe$_2$As$_2$ [48], and β-phase Mo$_{1-x}$Re$_x$ [49], where $\frac{\Delta T_c}{T_c}$ is broadening on the increase in $\frac{B_{appl}}{B_{c2}(0)}$ [42-49].

Here we report results of thorough analysis of $R(B,T)$ data for high-entropy alloy (ScZrNb)$_{0.65}$[RhPd]$_{0.35}$, and hydrogen-rich compounds of *Im-3m*-H$_3$S, *C2/m*-LaH$_{10}$ and *P6$_3$/mmc*-CeH$_9$. To perform the analysis we propose a new model to fit experimental $R(T,B)$ data, which also was successfully applied for $R(T,B=0)$ data of highly compressed elemental sulphur. We found that the dependence of $\frac{\Delta T_c}{T_c}$ vs $\frac{B_{appl}}{B_{c2}(0)}$ in $(ScZrNb)_{0.65}[RhPd]_{0.35}$, *Im-3m*-$H_3S$, *C2/m*-$LaH_{10}$ and *P6₃/mmc*-$CeH_9$ does follow a conventional trend of resistive transition broadening on the increase in applied magnetic field.

Experimental $R(T)$ data for *R3m*-phase of sulphur hydride and *Im-3m*-phase of sulphur deuteride was kindly provided by Dr. M. Einaga (Osaka University, Japan), $R(T)$ data for *Fm-3m*-phase of lanthanum hydride was kindly provided by Dr. M. I. Eremets and Dr. V. S. Minkov (Max-Planck Institut für Chemie, Mainz, Germany), and $R(T,B)$ data for *Im-3m*-phase of $H_3S$ was kindly placed [50] as open dataset by Dr. S. Mozafarri and co-authors (National High Magnetic Field Laboratory, Florida State University, USA).

**II. Model description**

Here we propose a single equation, which describes the full $R(T,B)$ curve, including the normal state part which is well above the onset of the resistive transition, $T \gg T_c^{onset}$, and the transition part, $T \lesssim T_c^{onset}$. To the best of the authors' knowledge, this sort of equations are not yet known, because existing models describe either the resistive part of $R(T,B)$ curves [51-56], or $R(T,B)$ near the $T_c^{onset}$ [41,57-59].

Our model is built on recent results [55,60], revealing that the normal part of $R(T,B=0)$ curves for a range of NRTS can be fitted to the Bloch-Grüneisen (BG) equation [56]:

$$R(T, B = 0) = R_0 + A \cdot \left(\frac{T}{T_\theta}\right)^5 \cdot \int_0^{\frac{T_\theta}{T}} \frac{x^5}{(e^x - 1) \cdot (1 - e^{-x})} \cdot dx \qquad (1)$$

where $R_0$, $A$ and $T_\theta$ are free-fitting parameters, and the latter is the Debye temperature. The deduced Debye temperature, $T_\theta$, and the observed transition temperature, $T_c$, are linked through the McMillan equation [61], which can be represented in the following advanced form [55]:

$$T_c = \left(\frac{1}{1.45}\right) \cdot T_\theta \cdot e^{-\left(\frac{1.04 \cdot (1+\lambda_{e-ph})}{\lambda_{e-ph} - \mu^* \cdot (1+0.62 \cdot \lambda_{e-ph})}\right)} \cdot f_1 \cdot f_2^* \qquad (2)$$

$$f_1 = \left(1 + \left(\frac{\lambda_{e-ph}}{2.46 \cdot (1+3.8 \cdot \mu^*)}\right)^{3/2}\right)^{1/3} \qquad (3)$$

$$f_2^* = 1 + (0.0241 - 0.0735 \cdot \mu^*) \cdot \lambda_{e-ph}^2. \qquad (4)$$

where $\mu^*$ is the Coulomb pseudopotential parameter (ranging from $\mu^* = 0.10$-$0.16$ [16,17,23, 32-40]), and $\lambda_{e-ph}$ is the electron-phonon coupling constant.

The system of equations Eqs. 2-4 has a unique solution in respect of $\lambda_{e-ph}$, if $T_\theta$, $T_c$ and $\mu^*$ are known. For the latter, in most cases, the mean value of $\mu^* = 0.13$ can be a good approximation, if $\mu^*$ was not computed for the given material by first principles calculations. As a result, the electron-phonon coupling constant, $\lambda_{e-ph}$, can be found by manual calculations due to substitution of relevant values in Eqs. 2-4.

There is a need for two clarifications associated with the approach (Eqs. 1-4):

1. The absence of the criterion to define the lower temperature limit for which $R(T,B=0)$ dataset should be fit to Eq. 1.

2. The absence of the criterion to define the transition temperature, $T_c$, from the curve $R(T,B=0)$ which can be used to calculate the electron-phonon coupling constant, $\lambda_{e\text{-ph}}$, by the use of Eqs. 2-4.

The latter problem has general implication, because it is equally applied on the results of first-principle calculations, where $T_c$ is one of the outcome parameters. However, there is no clarity which point in the experimentally recorded $R(T)$ curve corresponds to the computed $T_c$, because it can represent the temperature at the onset of the transition, $T_c^{onset}$, or zero resistance point, $T_{c,zero}$, or any temperature between these two experimental values, because $T_c$ can be defined by any ratio in the range:

$$0 \leq \left(\frac{R(T)}{R(T_c^{onset})}\right) \leq 1 \qquad (5)$$

(detailed discussion of the problem can be found elsewhere [55]).

Thus, the task is to find a function which reasonably well approximates the resistive transition, $R(T)$, and simultaneously self-stitches with the BG function (Eq. 1) at the onset of the superconducting transition, $T_c^{onset}$.

By experimenting with several functions which potentially can approximate the resistive transition and smoothly stitch the BG function, we report herein the result for a function which is similar, but not exact for a function proposed by Tinkham [59]:

$$R(T,B) = \frac{R_{norm}}{\left(I_0\left(\frac{C \cdot \left(1-\frac{T}{T_c}\right)^{3/2}}{2 \cdot B_{appl}}\right)\right)^2} \qquad (6)$$

where $I_0(x)$ is the zero-order modified Bessel function of the first kind and $C$ is a free-fitting parameter of Tesla units. This function was recently modified by Hirsch and Marsiglio [41]:

$$R(T,B) = \frac{R_{norm}}{\left(I_0\left(\frac{D \cdot \left(1-\frac{T}{T_c}\right)^{3/2}}{2 \cdot \frac{B_{appl}}{B_{c2}(0)}}\right)\right)^2} \qquad (7)$$

where $D$ is a dimensionless free-fitting parameter.

It should be stressed that there are several unavoidable problems associated with Eqs. 6,7. First of all, we can mention that a pivotal resistance curve $R(T, B_{appl} = 0)$ cannot be fitted to these equations, because the division by zero is prohibited. Also, as this was stated by Tinkham [59], $T_c$ in Eq. 6 (and in Eq. 7) is independent from the applied magnetic field which is (from his point of view [59]) an unphysical assumption of the model.

Here we propose a simpler function:

$$R(T,B) = \frac{R_{norm}}{\left(I_0\left(F \cdot \left(1-\frac{T}{T_c}\right)^{3/2}\right)\right)^2} \qquad (8)$$

which has three free-fitting parameters, $R_{norm}$, $T_c$ and $F$. Primary rational to use Eq. 8 is that multiplicative pre-factors:

$$\frac{1}{2 \cdot B_{appl}} \quad \text{in Eq. 6} \tag{9}$$

and

$$\frac{1}{2 \cdot \frac{B_{appl}}{B_{c2}(0)}} \quad \text{in Eq. 7} \tag{10}$$

only renormalize the value of free-fitting parameter $C$ and $D$ in Eqs. 6 and 7, respectively. As a result, Eqs. 6-8 provide essentially the same fits for $R(T, B \neq 0)$ datasets, however, Eq. 8 does not diverge at $B \to 0$.

We also need to stress that neither Tinkham [59], nor Hirsch and Marsiglio [41] propose any physical interpretation for the parameters $C$ and $D$ in their equations (Eqs. 6,7 respectively), and thus Eq. 8 can be derived following the same approach as Eqs. 6,7.

In result, the full equation which we propose to fit $R(T,B)$ curves in materials where charge carrier scattering on phonon is the dominant dissipation mechanism in normal state can be expressed in the following form:

$$R(T,B) = R_0 + k \cdot T + \theta(T_c^{onset} - T) \cdot \left( \frac{R_{norm}}{\left(I_0\left(F \cdot \left(1 - \frac{T}{T_c^{onset}}\right)^{3/2}\right)\right)^2} \right) + \theta(T - T_c^{onset}) \cdot$$

$$\left( R_{norm} + A \cdot \left( \left(\frac{T}{T_\theta}\right)^5 \cdot \int_0^{\frac{T_\theta}{T}} \frac{x^5}{(e^x-1) \cdot (1-e^{-x})} \cdot dx - \left(\frac{T_c^{onset}}{T_\theta}\right)^5 \cdot \int_0^{\frac{T_\theta}{T_c^{onset}}} \frac{x^5}{(e^x-1) \cdot (1-e^{-x})} \cdot dx \right) \right) \tag{11}$$

where $R_0$ and $k$ are free-fitting parameters which accommodate possible experimental onsets/uncertainties of the electronic measurement system, particular electrodes configuration in DAC, metallic weak links in the sample.

It should be noted that in order to be reliably fitted to Eq. 11, the normal part of the $R(T,B)$ curve should be measured in a reasonably wide temperature range, and, thus, if this is not the case (i.e., the measurements perform within narrow temperature range above $T_c^{onset}$), then more simple equation can be in use:

$$R(T,B) = R_0 + k \cdot T + \theta(T_c^{onset} - T) \cdot \left( \frac{R_{norm}}{\left( I_0 \left( F \cdot \left(1 - \frac{T}{T_c^{onset}}\right)^{3/2} \right) \right)^2} \right) +$$

$$\theta(T - T_c^{onset}) \cdot (R_{norm} + (k - k_1) \cdot T_c^{onset} + k_1 \cdot T) \qquad (12)$$

where $k_1$ is free-fitting parameter described a slope of normal part of $R(T,B)$ curve.

If the studied sample does not have metallic weak-links, then Eq. 12 is reduced to the equation with the same number of free-fitting parameters (which are $T_c^{onset}$, $R_{norm}$, $F$, $k_1$) as the standard fitting equation for the pining force density [62-65]:

$$F_p(B_{appl}) = F_{p,max} \cdot \left(\frac{B_{appl}}{B_{c2}}\right)^p \cdot \left(1 - \frac{B_{appl}}{B_{c2}}\right)^q \qquad (13)$$

where $F_{p,max}$, $B_{c2}$, $p$ and $q$ are free-fitting parameters. For these samples, full Eq. 11 exhibits 5 free-fitting parameters, where the additional parameter is the Debye temperature, $T_\theta$, which is a fundamental characteristic of the superconductor, and, thus, $T_\theta$ cannot be asserted to be an unnecessary parameter.

To fit the $R(T,B)$ datasets we use a non-linear data fit tool from the ORIGIN package (ver. 2017). In all plots we show 95% confidence bands for the fitted curves, which calculated by the ORIGIN package and these bands represent $2\sigma$ uncertainty for the fitted curve calculated from $2\sigma$ uncertainties of all free-fitted parameters in each point. Rigorous mathematical definition of 95% confidence bands can be found elsewhere [66].

From converged fit, $T_c$ can be defined by any chosen $\frac{R(T)}{R(T_c^{onset})}$ criterion (Eq. 5). In this work we use the value of $T_{c,0.03}$:

$$\frac{R(T)}{R(T_c^{onset})} = 0.03 \qquad (14)$$

which was used to calculate the electron-phonon coupling constant, $\lambda_{\text{e-ph}}$ (Eqs. 2-4). This value, $T_c^{0.03}$, was also used to deduce the width of the resistive transition, $\Delta T_c$:

$$\frac{\Delta T_c}{T_c} = \frac{T_c^{onset} - T_{c,0.03}}{T_c^{onset}} \qquad (15)$$

The reason for choosing $\frac{R(T)}{R(T_c^{onset})} = 0.03$ for the analysis of experimental $R(T,B)$ data of NRTS originates from a requirement, that this ratio should be as small as possible (extended discussion of this important issue is given in Ref. 55) on the one hand. On the other hand, this ratio should be well above the level of noise of experimental $R(T,B)$, as the analysis will be performed for real measured $R(T,B)$ data, and not for extrapolated tails of given mathematical functions. Based on available $R(T,B)$, the ratio of $\frac{R(T)}{R(T_c^{onset})} = 0.03$ (Eq. 13) fulfils both requirements.

### III. Results

#### 3.1. Highly-compressed superconductors in zero magnetic field

#### *Elemental sulphur*

Before the model (Eq. 11) will be applied for NRTS materials, we demonstrate its applicability for highly-compressed elemental sulphur (Fig. 1). Experimental $\frac{R(T)}{R(T=77K)}$ curves were reported by Shimizu *et al* [67] (in their Fig. 10 [67]).

The model perfectly found $T_c^{onset}$ for all $\frac{R(T)}{R(T=77K)}$ datasets (Fig. 1). All fits converged with excellent quality (where the goodness of fit, $R$, is varying within a narrow range of 0.997-0.998 for all fits in Fig. 1). Free-fitting parameters of the model $T_\theta$, as well as the calculated parameters $T_{c,0.03}$ and $\lambda_{e-ph}$ (for which we used Eqs. 2-4,14) are also smoothly varying within narrow ranges (Fig. 1). The superconducting transition width, $\frac{\Delta T_c}{T_c}$, has a trend to increase vs the increase in applied pressure.

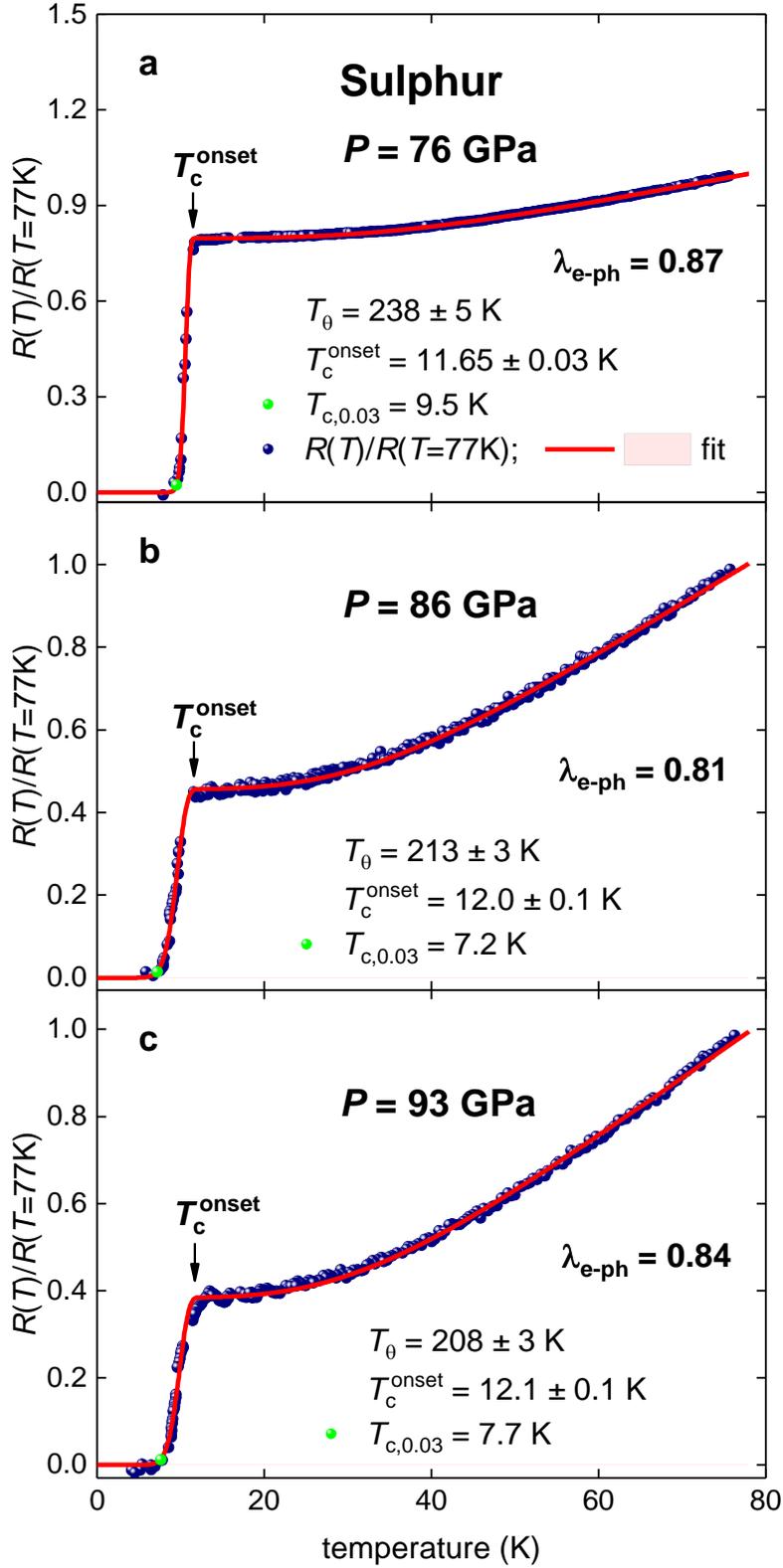

**Figure 1.** $R(T)/R(T=77K)$ data for highly-compressed sulphur (raw data is from Ref. 66) fitted to Eq. 11. Green balls show $T_{c,0.03}$. Red is the fitting curve, 95% confidence bars are shown for all panels by a pink shaded area, which is narrower than the thickness of the fitting curve. a – $\frac{\Delta T_c}{T_c} = 0.18$; b – $\frac{\Delta T_c}{T_c} = 0.40$; c – $\frac{\Delta T_c}{T_c} = 0.36$.

*R3m-phase of sulphur hydride*

Einaga *et al* [18] studied the pressure dependence of the transition temperature, $T_c(P)$, in *R3m*-phase of highly-compressed sulphur hydride. In Fig. 2 we show raw $R(T, P = 133$ GPa) curve for this phase (reported in Fig. 3(a) [18]), and data fit to Eq. 11.

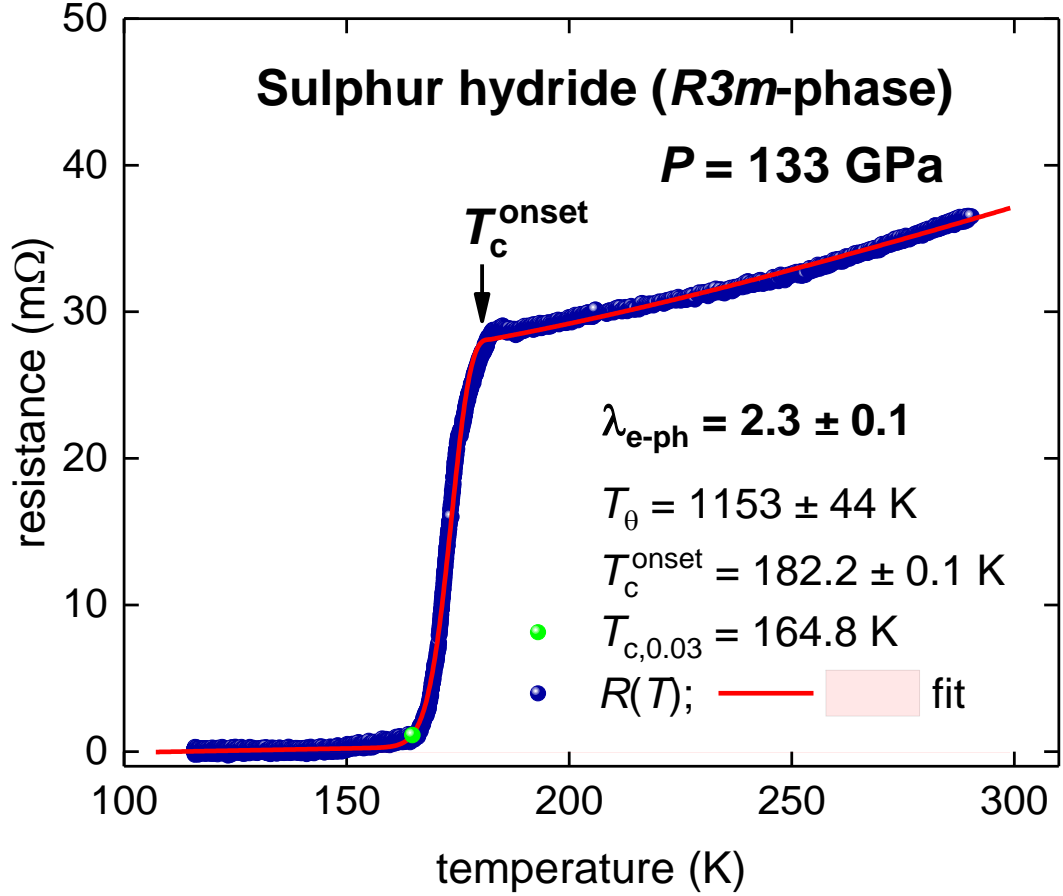

**Figure 2.** *R(T)* data for highly-compressed *R3m*-phase of sulphur hydride (raw data is from Ref. 18) and data fit to Eq. 11. Green ball shows $T_{c,0.03}$. 95% confidence bars are shown by a pink shaded area, which is narrower than the thickness of the fitting curve; goodness of fit is 0.9993.

Deduced $\lambda_{e-ph}(P = 133\ GPa) = 2.3 \pm 0.1$ is very close to the first principle calculations value of $\lambda_{e-ph}(P = 130\ GPa) = 2.07$ reported by Duan *et al* [68] in their first pivotal paper on highly-compressed H-S system. The superconducting transition has a moderate width of $\frac{\Delta T_c}{T_c} = 0.095$.

*Im-3m-phase of sulphur deuteride*

Einaga *et al* [18] also studied the pressure dependence of the transition temperature, $T_c(P)$, in the *Im-3m*-phase of sulphur deuteride. In Fig. 3 we show the raw $R(T, P = 190$ GPa) data for this phase (reported in Fig. 3(b) [18]) and data fit to Eq. 11.

The fit in Fig. 3 is a compelling example for the model validity, because it covers a wide temperature range of $14\ K\ \leq T \leq 290\ K$ with remarkably high quality of 0.99994.

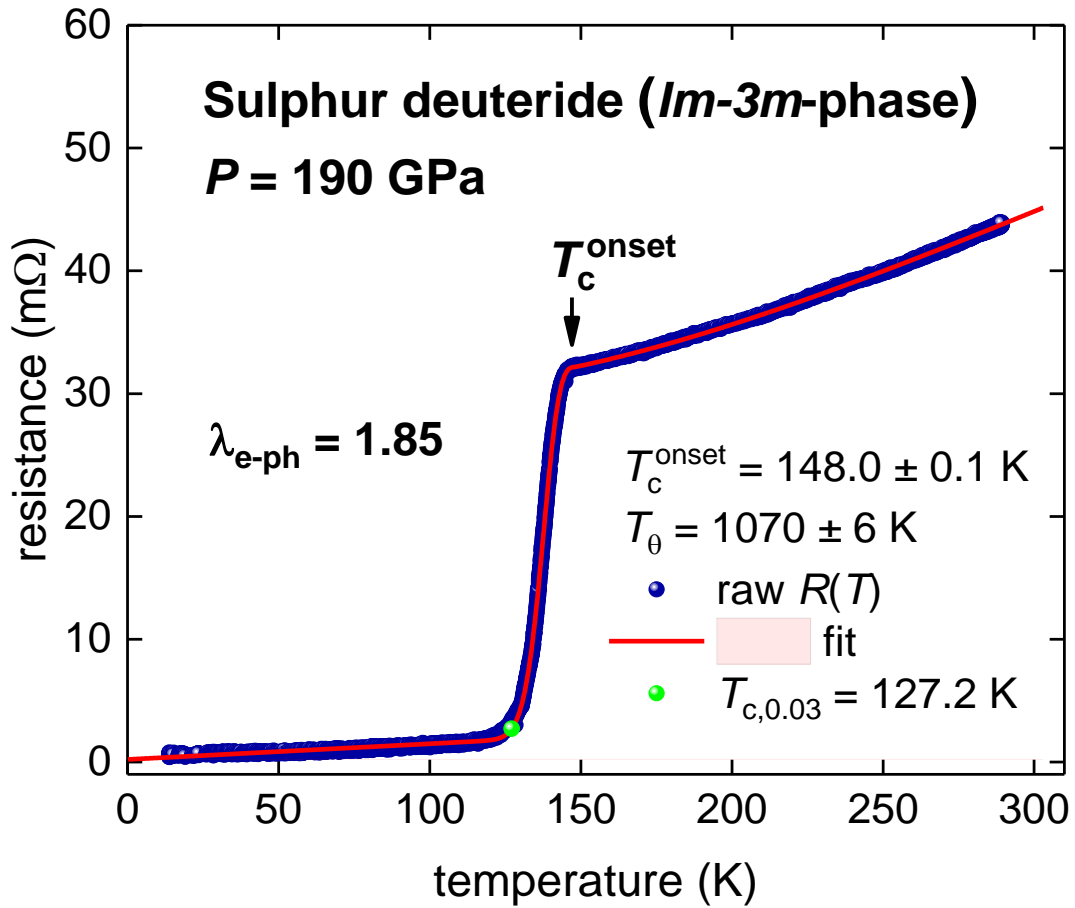

**Figure 3.** $R(T)$ data for highly-compressed *Im-3m*-phase of D$_3$S (raw data is from Ref. 18) and data fit to Eq. 11. Green ball shows $T_{c,0.03}$. 95% confidence bars are shown by a pink shaded area, which is narrower than the thickness of the fitting curve; goodness of fit is 0.99994.

Deduced $\lambda_{e-ph} = 1.85$ is also within expected range reported for this value by first principles calculations, $\lambda_{e-ph} = 1.86$ [36]), and the superconducting transition has moderate width of $\frac{\Delta T_c}{T_c} = 0.14$.

*Hydrogen deficient lanthanum hydride LaH$_x$ (x > 3)*

Somayazulu *et al* [25] and Drozdov *et al* [26] independently reported on the discovery of NRTS in several phases of the La-H system, from which the highest transition temperature of $T_c = 245 - 280\ K$ was observed in *Fm-3m*-phase of LaH$_{10}$ [26]. The transition temperature depends on the hydrogen stoichiometry and the *R(T)* curve for a highly-hydrogen deficient sample (Sample 11 [26]) is shown in Fig. 4 together with the *R(T)* fit to Eq. 11. The quality of the fit is excellent, and this sample has a moderately wide superconducting transition width of $\frac{\Delta T_c}{T_c} = 0.12$. Deduced the electron-phonon coupling constant, $\lambda_{e-ph} = 1.68$ (*P* = 150 GPa, *T*$_c$ = 73 K). It should be noted that the sample (Sample 11, LaH$_x$ (x>3), *P* = 150 GPa [26]) has a complex XRD pattern and the dominant phase in this sample is unknown [26]. This is the reason, why direct comparison of the deduced $\lambda_{e-ph} = 1.68$ (*P* = 150 GPa) value with its contemplates calculated for stoichiometric hydrogen-rich phases of LaH$_{10}$ [32,69] cannot be made at quantitative level, because stoichiometric phases have *T*$_c$ > 240 K vs *T*$_c$ = 73 K for Sample 11 (Fig. 4). However, there is a possibility for qualitative comparison. The lower limit for the electron-phonon coupling constant was reported by Durajski and Szczęśniak [70], who calculated *T*$_c$ = 22.5 K and $\lambda_{e-ph} = 0.845$ for LaH$_3$ compressed at *P* = 11 GPa. The same research group reported *T*$_c$ = 215 K and $\lambda_{e-ph} = 2.18$ for LaH$_{10}$ compressed at *P* = 150 GPa [69], which can be assumed to be the upper limiting values. Thus, deduced herein *T*$_c$ = 73 K and $\lambda_{e-ph} = 1.68$ for LaH$_x$ (x > 3) compressed at *P* = 150 GPa are in between of the lower and the upper limits [69,70].

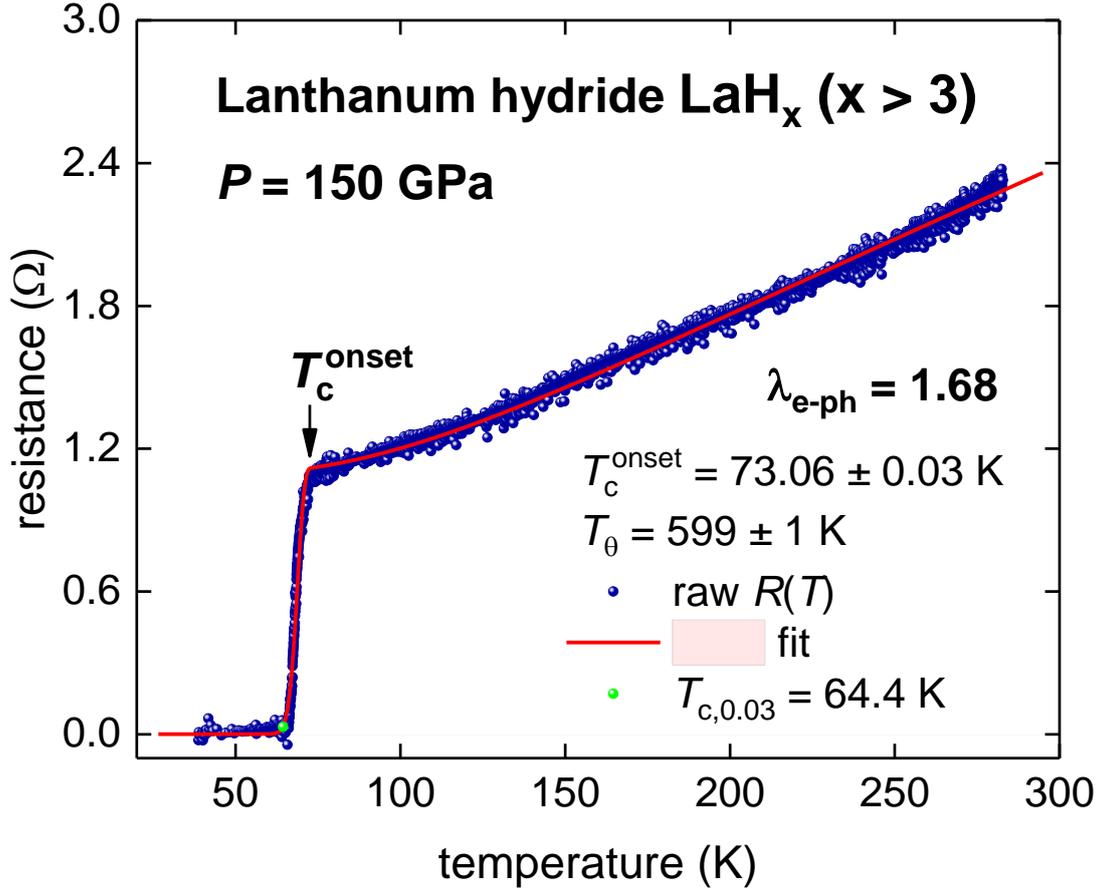

**Figure 4.** $R(T)$ data for highly-compressed hydrogen deficient LaH$_x$ (x > 3) (raw data is from Sample 11 [23]) and data fit to Eq. 11. Green ball shows $T_{c,0.03}$. 95% confidence bars are shown by a pink shaded area, which is narrower than the thickness of the fitting curve; goodness of fit is 0.997.

*Fm-3m-phase of lanthanum hydride*

Drozdov *et al* [26] also reported $R(T)$ data for stoichiometric *Fm-3m*-phase of LaH$_{10}$ (Sample 1 [26]) which is shown in Fig. 5. Due to the normal part of $R(T)$ curve was measured at narrow temperature range, the fit was performed to Eq. 12 (Fig. 5). Despite a high transition temperature, $T_{c,0.03} = 236\ K$, this sample has a narrow transition width of $\frac{\Delta T_c}{T_c} = 0.056$.

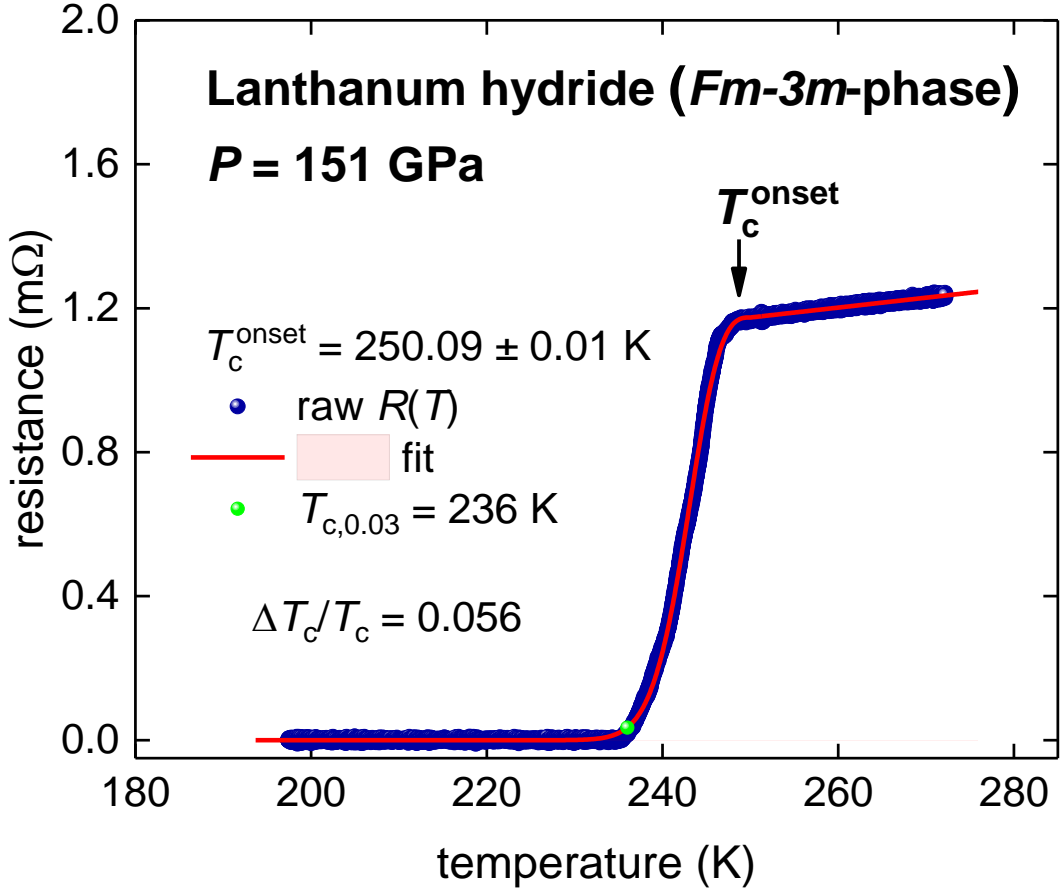

**Figure 5.** *R*(*T*) data for highly-compressed *Fm-3m*-phase of LaH$_{10}$ (Sample 1 [23]) and data fit to Eq. 12. Green ball shows $T_{c,0.03}$. 95% confidence bars are shown by a pink shaded area, which is narrower than the thickness of the fitting curve; goodness of fit is 0.9998.

*3.2. Superconductors in applied magnetic field*

*High-entropy alloy (ScZrNb)$_{0.65}$[RhPd]$_{0.35}$ in applied magnetic field*

To extend the dependence of $\frac{\Delta T_c}{T_c}$ vs $\frac{B_{appl}}{B_{c2}(0)}$ [41] for a wider class of materials, we applied our model for *R*(*T,B*) measured in high-entropy alloy (ScZrNb)$_{0.65}$[RhPd]$_{0.35}$ [71]. Details of the experiment were reported elsewhere [71] and here in Fig. 6 we show the *R*(*T,B*=0) data fit to Eq. 11. The fit is excellent and the deduced $\lambda_{e\text{-ph}}$ = 1.03 is close to recently reported $\lambda_{e\text{-ph}}$ = 1.10 for another high-entropy alloy (TaNb)$_{0.67}$(HfZrTi)$_{0.33}$ [72].

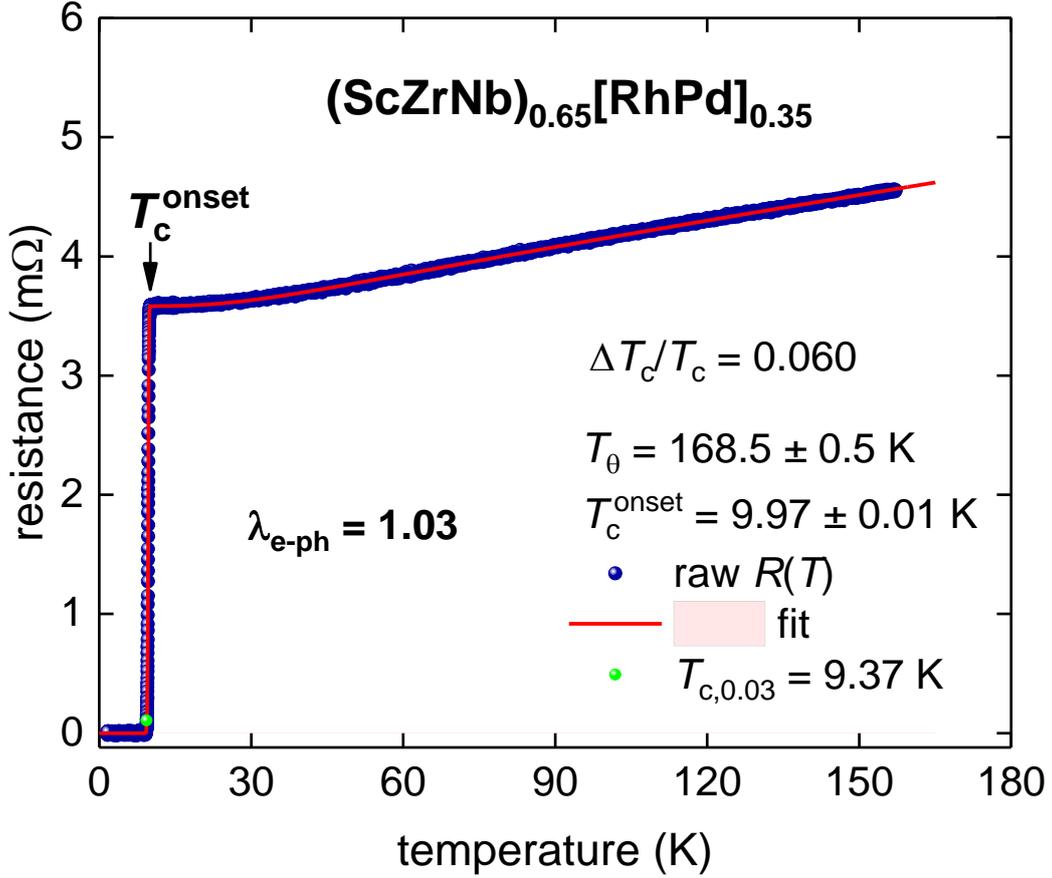

**Figure 6.** $R(T,B=0)$ data for high-entropy alloy $(ScZrNb)_{0.65}[RhPd]_{0.35}$ and data fit to Eq. 11. Green ball shows $T_{c,0.03}$ defined by $R(T)/R_{norm}(T) = 0.03$ criterion. 95% confidence bars are shown by a pink shaded area, which is narrower than the thickness of the fitting curve; goodness of fit is 0.99997.

As the $R(T,B)$ measurements were performed in a more narrow temperature range of $1.7\ K \leq T \leq 12\ K$, we fit these datasets to Eq. 12. All fits have excellent quality (with good of fitness > 0.99990) and some of these fits are shown in Fig. 7. In the result, the plot of $\frac{\Delta T_c}{T_c}$ vs $\frac{B_{appl}}{B_{c2}(0)}$, with $B_{c2}(0) = 10.7$ T [71], is shown in Fig. 8, where the conventional trend of the transition width broadening on applied magnetic field is apparent.

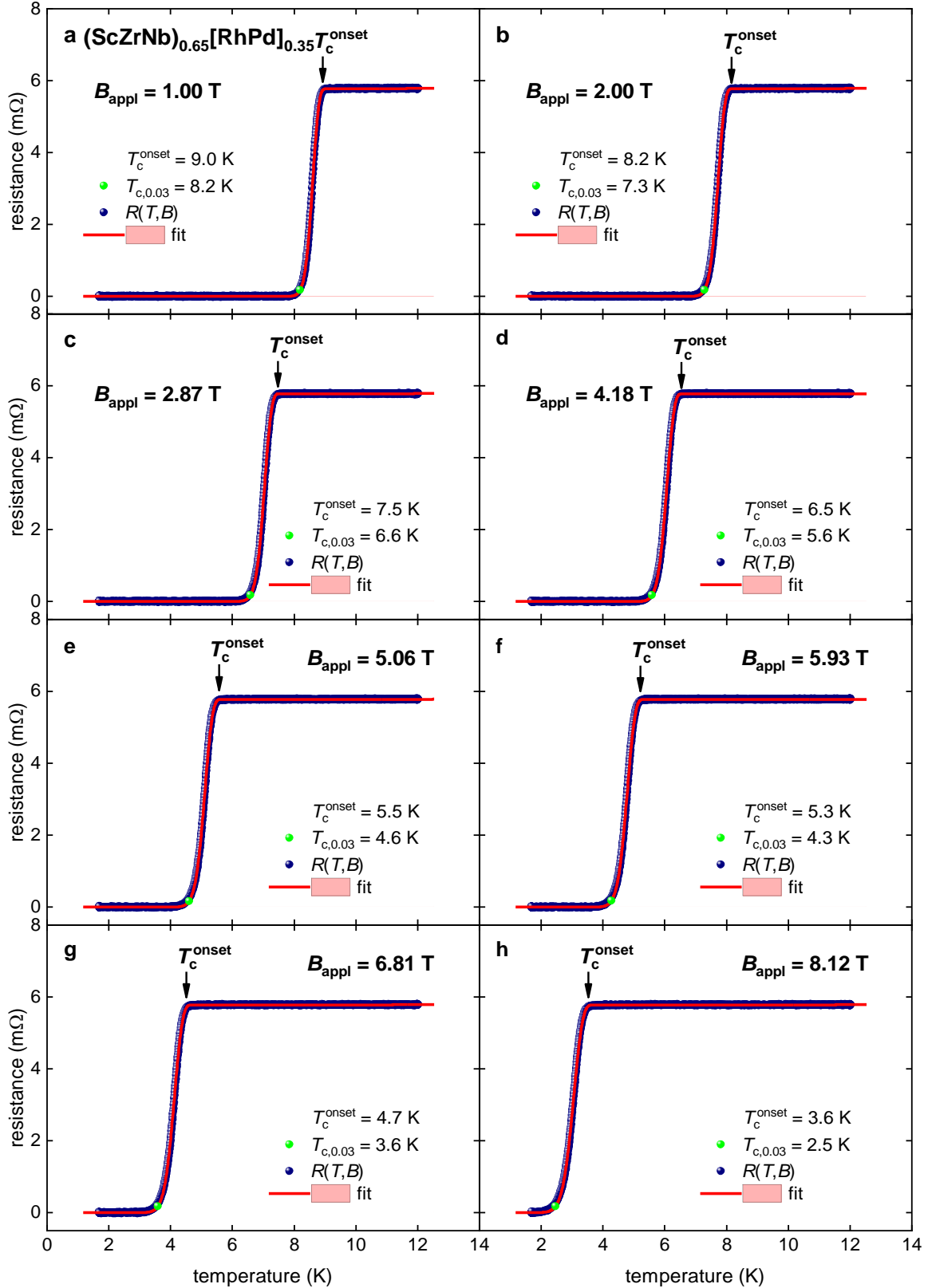

**Figure 7.** $R(T,B)$ data for high-entropy alloy $(ScZrNb)_{0.65}[RhPd]_{0.35}$ and data fit to Eq. 12. Goodness of fit for all fits is better than 0.99990. 95% confidence bars are shown by a pink shaded area, which is narrower than the thickness of the fitting curve.

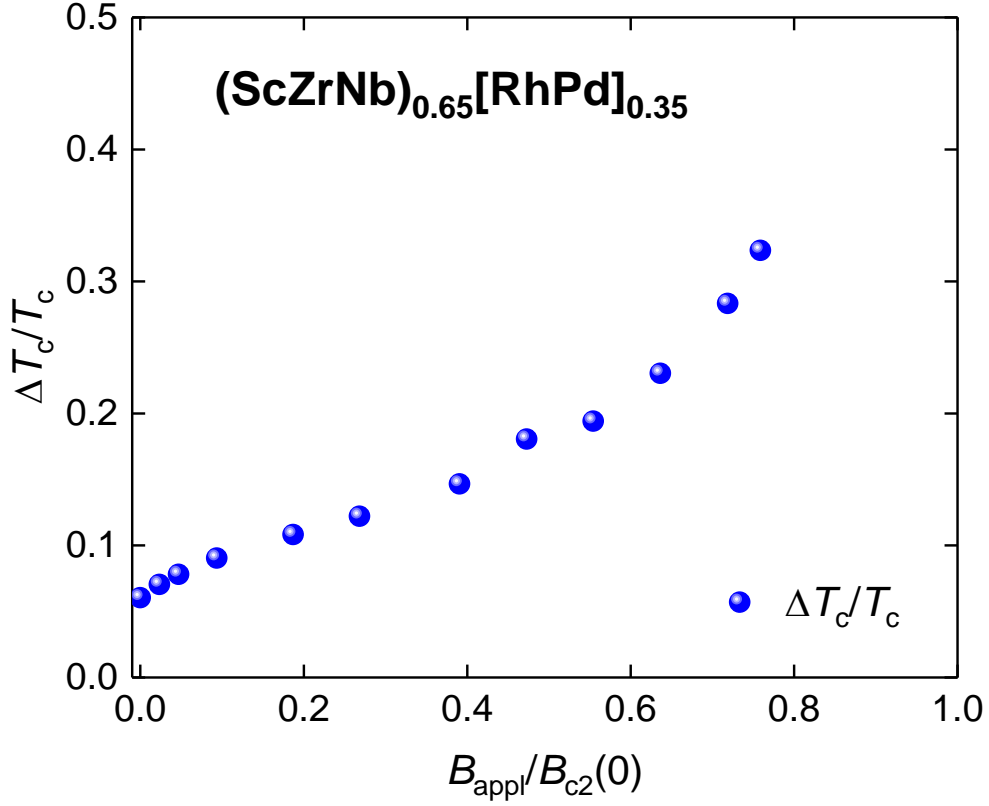

**Figure 8.** Magnetic field dependence of the reduced superconducting transition width, $\Delta T_c/T_c$, for high-entropy alloy $(ScZrNb)_{0.65}[RhPd]_{0.35}$.

### *Im-3m-phase of H₃S in applied magnetic field*

Mozaffari *et al* [50] reported high-field magnetoresistance measurements, $R(T,B)$, for *Im-3m*-phase of $H_3S$ compressed at $P = 155$ GPa and 160 GPa. Here we analyse data for the sample compressed at $P = 155$ GPa which exhibits two resistive transitions at zero applied field (Fig. 9). The $R(T)$ data was fitted to a two-step function:

$$R(T,B) = R_0 + \theta(T_{c1}^{onset} - T) \cdot \left(\frac{R_{norm,1}}{\left(I_0\left(F_1 \cdot \left(1 - \frac{T}{T_{c1}^{onset}}\right)^{3/2}\right)\right)^2}\right) + \theta(T_{c2}^{onset} - T) \cdot \theta(T - T_{c1}^{onset}) \cdot$$

$$\left(\frac{R_{norm,2}}{\left(I_0\left(F_2 \cdot \left(1 - \frac{T}{T_{c2}^{onset}}\right)^{3/2}\right)\right)^2}\right) + \theta(T - T_{c2}^{onset}) \cdot (R_{norm,1} + R_{norm,2} + k \cdot T) \qquad (16)$$

where subscripts 1 and 2 indicate the transition.

The fit reveals that both transitions have very narrow transition widths of $\frac{\Delta T_{c1}}{T_{c1}} = 0.009$ and $\frac{\Delta T_{c2}}{T_{c2}} = 0.007$.

In Fig. 10 we fit to Eq. 12 (where *k* was fixed to zero) four *R*(*T*,*B*) datasets of *Im-3m*-phase of H$_3$S (*P* = 155 GPa) reported by Mozaffari *et al* [50] in their Supplementary Figures 1 and 2. An important feature of the result is that fits of *R*(*T*,*B*) datasets for *Im-3m*-phase of H$_3$S (*P* = 155 GPa) were processed by the same mathematical routine and the same criterion was applied to deduce $\frac{\Delta T_c}{T_c}$ vs $\frac{B_{appl}}{B_{c2}(0)}$ dependence. Based on this, the result is not distorted by any variation, which can be appeared if manual data processing would be implemented.

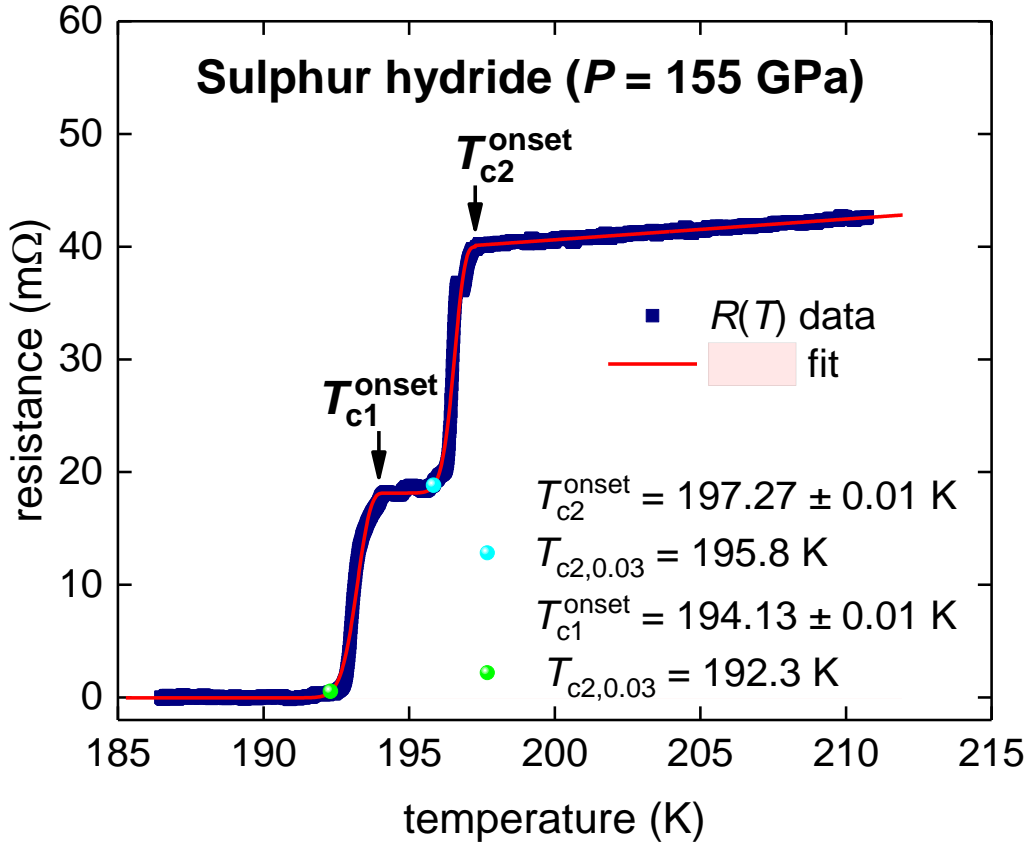

**Figure 9.** *R*(*T*,*B*=0) data for highly-compressed *Im-3m*-phase of H$_3$S (*P* = 155 GPa [50]) and data fit to Eq. 16. Green and cyan balls show $T_{c,0.03}$ defined by $R(T)/R_{norm}(T) = 0.03$ criterion for both transitions. 95% confidence bars are shown a pink shaded area, which is narrower than the thickness of the fitting curve; goodness of fit is 0.9992.

It can be seen in Fig. 11 that $\frac{\Delta T_c}{T_c}$ vs $\frac{B_{appl}}{B_{c2}(0)}$ (where we adopt $B_{c2}(0)$ = 88 T reported by Mozaffari *et al* [50] for this sample), for *Im-3m*-phase of H$_3$S does follow general trend of the transition width broadening on the increase of the magnetic field. For instance, $\frac{\Delta T_c}{T_c}$ vs $\frac{B_{appl}}{B_{c2}(0)}$ curve for H$_3$S has similar trend with ones for Nb$_3$Sn and BaFe$_{2-x}$Ru$_x$As$_2$ (x = 0.71) (Fig. 11).

*C2/m-phase of LaH$_{10}$ in applied magnetic field*

Sun *et al* [29] reported on the discovery of new *C2/m*-phase of LaH$_{10}$ compound, which exhibits zero resistance at T = 170-185 K at pressure range of *P* = 120-130 GPa. Sun *et al* [29] also reported high-field magnetoresistance measurements, *R(T,B)*, for this *C2/m*-phase compressed at *P* = 120 GPa, which we analysed in our Fig. 12.

It can be seen in Fig. 11 that the $\frac{\Delta T_c}{T_c}$ vs $\frac{B_{appl}}{B_{c2}(0)}$ dependence (for which we adopt $B_{c2}(0)$ = 133.5 T reported by Sun *et al* [29] for this phase) follows the general trend of the transition width broadening with increased applied magnetic field.

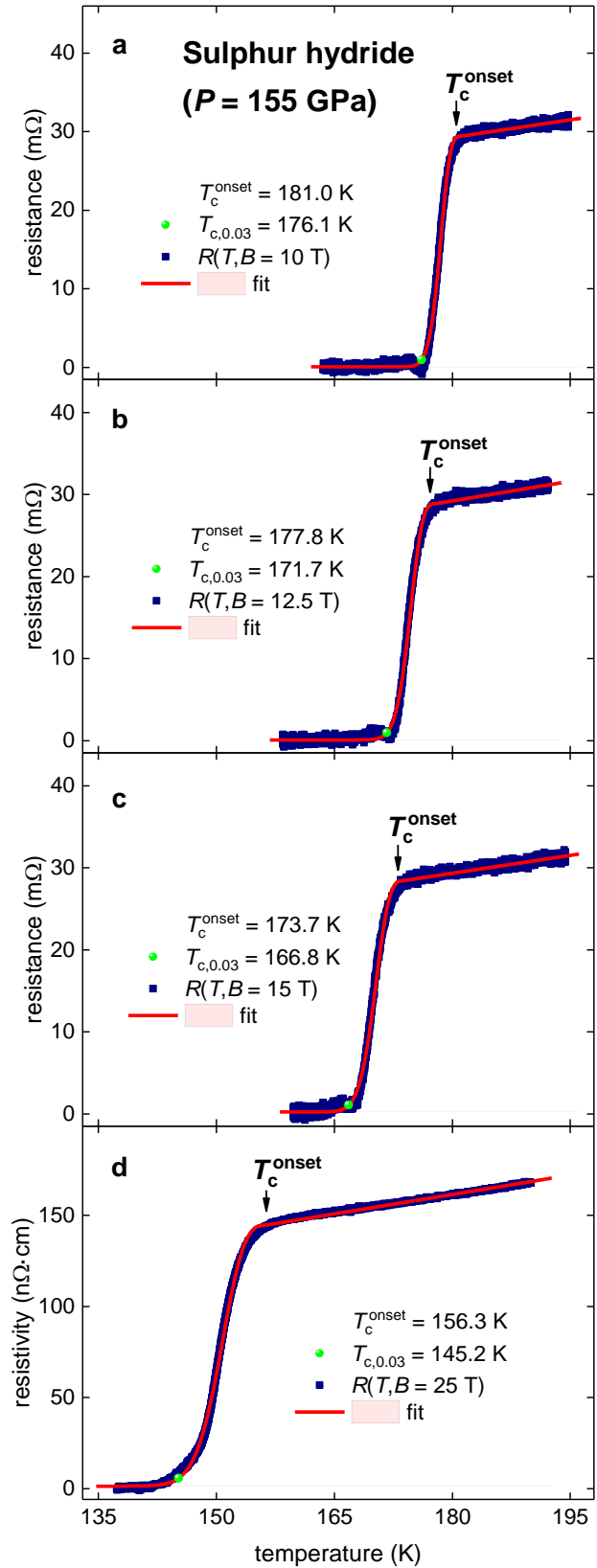

**Figure 10.** $R(T,B)$ data for highly-compressed *Im-3m*-phase of $H_3S$ ($P$ = 155 GPa) and data fit to Eq. 12 ($k$ was fixed to zero). Green balls show $T_{c,0.03}$. 95% confidence bars are shown by a pink shaded area, which is narrower than the thickness of the fitting curve, goodness of fit for all curves is better than 0.9991.

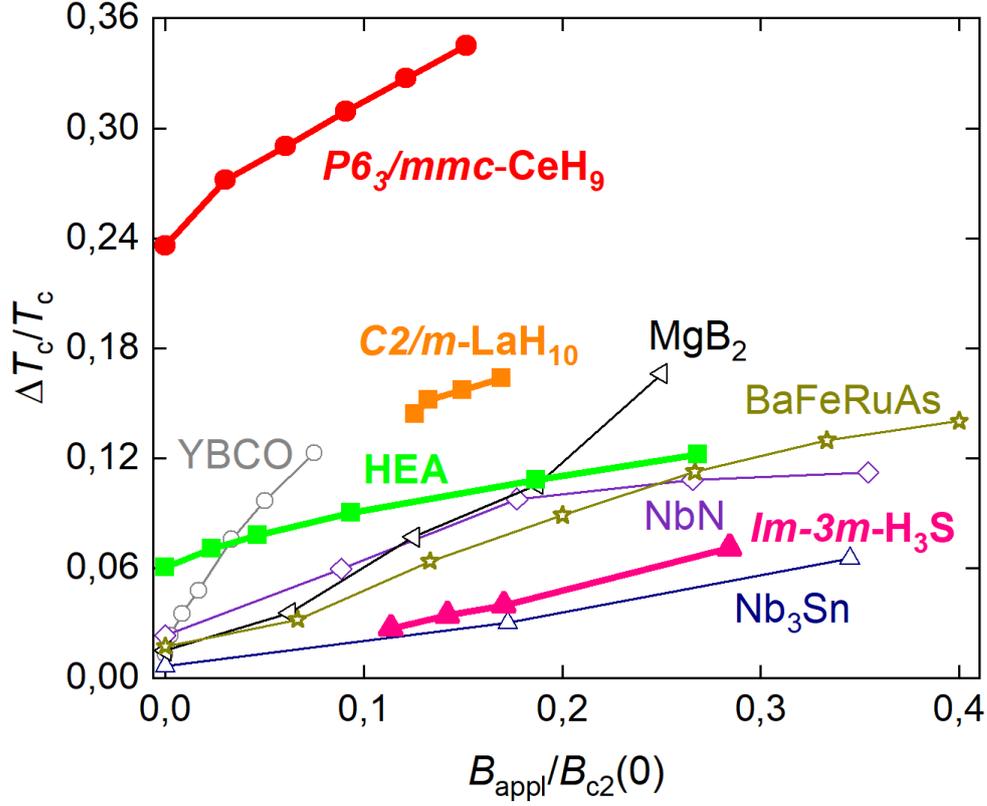

**Figure 11.** Field dependence of the superconducting transition width, $\Delta T_c/T_c$, vs reduced applied field, $B_{appl}/B_{c2}(0)$, for materials processed in this paper (depicted in bold) and several representative materials of major superconducting families (data for these materials was adopted from Ref. 41).

*P6₃/mmc-phase of CeH₉ in applied magnetic field*

Chen *et al* [16] synthesized a new high-temperature superconducting *P6₃/mmc*-phase of CeH$_9$ which exhibits $T_c$ = 45-85 K at relatively low pressure of $P$ = 88-140 GPa. Chen *et al* [16] reported extensive magnetoresistive studies of this new phase, from which we analysed Sample Cell #H1 compressed at $P$ = 88 GPa. The raw $R(T,B)$ data is reported in Fig. S7 [16].

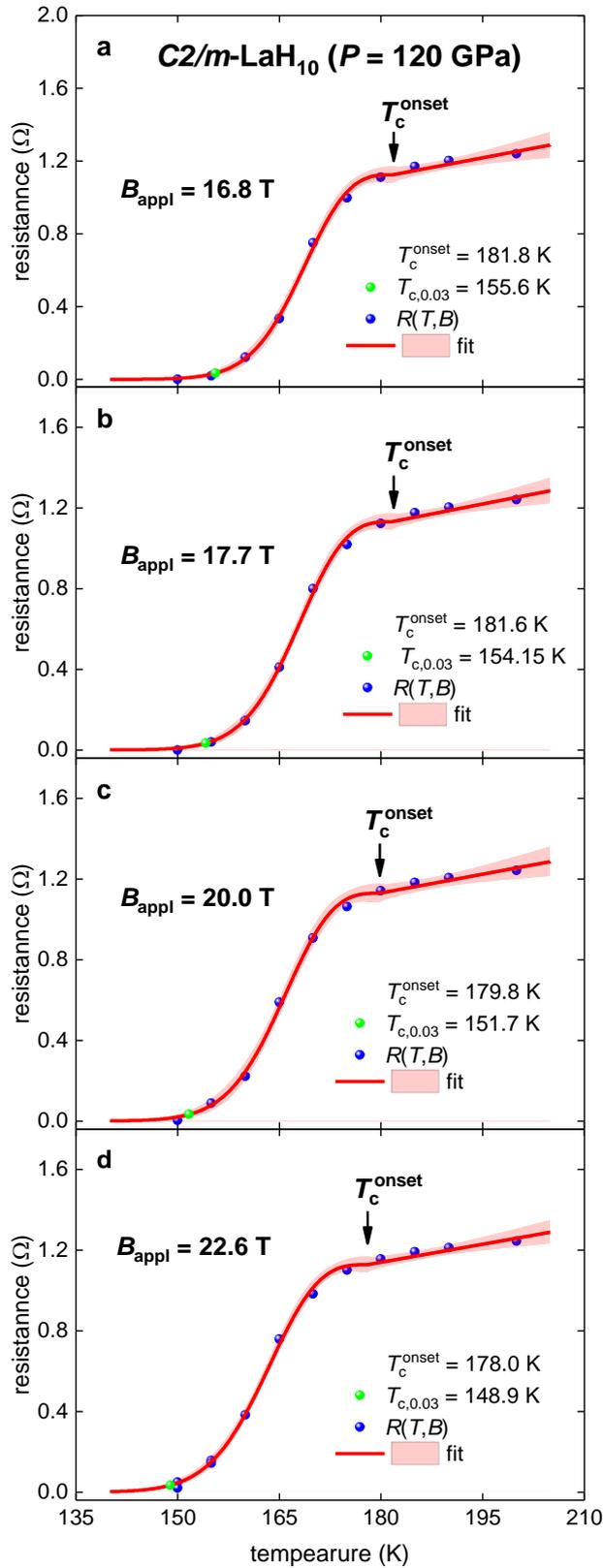

**Figure 12.** $R(T,B)$ data for highly-compressed $C2/m$-phase of LaH$_{10}$ ($P$ = 120 GPa) and data fit to Eq. 12 ($k$ and $R_0$ were fixed to zero). Green balls show $T_{c,0.03}$. Red is fitting curve, 95% confidence bars are shown by a pink shaded area, goodness of fit for all curves is better than 0.998.

All fits to Eq. 12 for this sample are shown in Fig. 13 and all fits have excellent quality. Deduced $\frac{\Delta T_c}{T_c}$ vs $\frac{B_{appl}}{B_{c2}(0)}$ dependence (for which we adopt $B_{c2}(0) = 33$ T reported by Chen *et al* [16] for this sample), are shown in Fig. 11, and the usual trend of the transition with broadening vs the increase in applied magnetic field is obvious (Fig. 11).

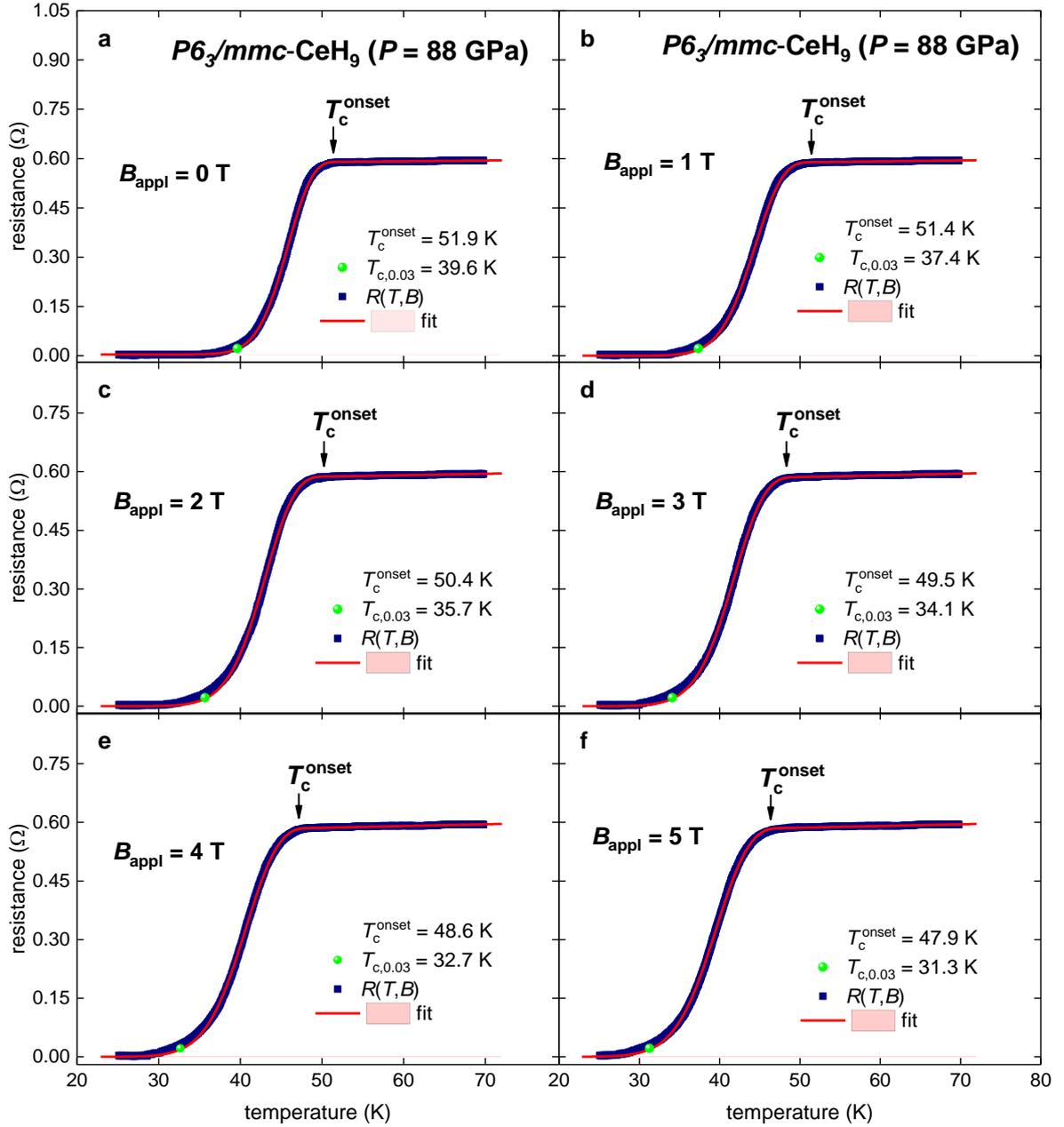

**Figure 13.** $R(T,B)$ data for highly-compressed $P6_3/mmc$-phase of CeH$_9$ ($P = 88$ GPa) and data fit to Eq. 12 ($k$ and $R_0$ were fixed to zero). Green balls show $T_{c,0.03}$. 95% confidence bars are shown by a pink shaded area, which is narrower than the thickness of the fitting curve, goodness of fit for all curves is better than 0.9994.

## IV. Discussion

We show above that hydrogen-rich superconductors exhibit a usual trend of the broadening of the resistive transition width, $\frac{\Delta T_c}{T_c}$, on the increase in applied magnetic field, $\frac{B_{appl}}{B_{c2}(0)}$. This result is in contrast with a recent report by Hirsch and Marsiglio [41], who found that the transition width, $\frac{\Delta T_c}{T_c}$, remains constant (or even supresses in case of YH$_9$) on the increase of $\frac{B_{appl}}{B_{c2}(0)}$.

To explain this apparent contradiction, we need to point out several important issues. First of all, Hirsch and Marsiglio [41] analysed $R(T,B)$ data measured at low applied magnetic fields:

$$\frac{B_{appl}}{B_{c2}(0)} \leq 0.1 \tag{16}$$

where $B_{c2}(0)$ was obtained by an extrapolation of $B_{c2}(T)$ data also measured at a narrow temperature range near $T_c$:

$$0.9 \leq \frac{T}{T_c} \leq 1.0 \tag{17}$$

which both create reasonable uncertainty in the analysis [41].

However, more important is the fact that from our best knowledge, Hirsch and Marsiglio [41] did not provide a $\Delta T_c$ definition used for their analysis, and thus, there is no clarity on how $\frac{\Delta T_c}{T_c}$ was deduced from the experimental data.

In addition, the NRTS materials analysed in Ref. 41 were multiphase samples and if these phases have close transition temperatures at $B = 0$ T (see, for instance, Fig. 9) then this can lead to an effective broadening of the total resistive transition $R(T,B = 0)$. However, different superconducting phases are more likely to have reasonably different upper critical fields, $B_{c2}(T)$. And by applying a magnetic field, $B_{appl}$, the superconducting state in one (or several) phase can be completely suppressed. Thus, at some applied field the superconducting state will remain in one phase, which exhibits the highest upper critical field. As a result, the transport current will be short circuited within the remaining superconducting phase, which will be effectively appearing as a "narrowing" of the in-field resistive transition, $R(T,B)$. This effect of extinction of superconducting phases is exhibited at relatively low applied magnetic field. It can clearly be seen in case of $H_3S$ in Supplementary Figure 1 [50]. Because secondary phases are eliminated at low applied magnetic fields near $T_c$, at higher applied fields the resistance curve, $R(T,B)$, is solely dependent on a single superconducting phase and the resistive transition appears to be sharper, than in the case where several phases contribute to $R(T,B)$. However, at higher applied magnetic field and respectively lower temperatures conventional resistive transition broadening vs applied magnetic field is expected, because it is related to a single superconducting phase. And this is what we find in our report herein (Fig. 11), where $R(T,B)$ data measured at $0.1 \leq \frac{B_{appl}}{B_{c2}(0)} \leq 0.3$ in case of $H_3S$ and $LaH_{10}$ were analysed, and for the single-phase $P6_3/mmc$-$CeH_9$ sample for which measurements were performed in the range of $0.0 \leq \frac{B_{appl}}{B_{c2}(0)} \leq 0.15$. The latter case confirms that the broadening at low applied fields is still valid for hydrogen-rich superconductors, if the sample is a single phase.

Our explanation received independent confirmation as a result of thorough literature search. For instance, Wang *et al* [73] showed that the transition width, $\Delta T_c$, in SiC-doped polycrystalline $MgB_2$ samples decreases in weak magnetic fields. At some medium fields, $\Delta T_c(B)$ reaches its minimum and starts to linearly increase at higher magnetic fields. This is a compelling example of non-monotonic $\Delta T_c(B)$ in multiphase polycrystalline samples, which were all samples studied in Ref. 41.

It is important to discuss possible limits for the applicability of our basic Eqs. 11, 12 to describe the resistive transition, $R(T,B)$. From its definition, Eq. 11 is applicable for materials where charge carriers predominantly scatter by the lattice vibrations (i.e., phonons). However, for materials where the observed $T_c$ does not solely depend on the electron-phonon interaction (particularly, in cuprates [51,53,54,], pnictides [74-76], nickelates [77,78]), but instead depends on the thermodynamic fluctuations [58,79-82], the presence of pseudogap [83,84], structural disorder [85-88], system dimensionality [89-93] or other mechanisms [94-97] for charge carriers wave function distortions, our Eqs. 11,12 might not be a good fitting tool (however, it does not necessarily mean that in some case these equations will be still a good fitting tool). Based on all above, alternative $R(T,B)$ models may be developed for other types of materials.

## V. Conclusion

In this work we study the in-field dependence of the reduced resistive transition width, $\frac{\Delta T_c}{T_c}$, in the high-entropy alloy $(ScZrNb)_{0.65}[RhPd]_{0.35}$ and hydrogen-rich superconductors *Im-3m*-$H_3S$, *C2/m*-$LaH_{10}$ and *P6$_3$/mmc*-$CeH_9$. To perform the analysis, we propose a new function to fit $R(T,B)$ curves and a strict mathematical routine to deduce $\Delta T_c$ from the analysis of $R(T,B)$ curves. As a result, we show that the reduced transition width, $\frac{\Delta T_c}{T_c}$, in $(ScZrNb)_{0.65}[RhPd]_{0.35}$, *Im-3m*-$H_3S$, *C2/m*-$LaH_{10}$ and *P6$_3$/mmc*-$CeH_9$ does follow a conventional broadening trend with increased applied magnetic field.

**Acknowledgement**

The authors thank Dr. M. Einaga (Osaka University, Japan) for providing experimental data for *R3m*-phase of sulphur hydride and *Im-3m*-phase of sulphur deuteride, Dr. M. I. Eremets and Dr. V. S. Minkov (Max-Planck Institut für Chemie, Mainz, Germany) for providing data for *Im-3m*-phase of sulphur hydride and *Fm-3m*-phase of lanthanum hydride, Dr. S. Mozafarri and co-authors (National High Magnetic Field Laboratory, Florida State

University, USA) for open access magnetoresistance data for *Im-3m*-phase of $H_3S$ [50], and P. A. Provencher (Princeton University, USA) for proofreading the manuscript.

EFT thanks financial support provided by the Ministry of Science and Higher Education of Russia (theme "Pressure" No. AAAA-A18-118020190104-3) and by Act 211 Government of the Russian Federation, contract No. 02.A03.21.0006.

**Author Contributions**

EFT conceived the idea and proposed Eqs. 11,12,16. KS analysed data for high-entropy alloy and EFT analysed data for highly-compressed superconductors. EFT wrote the manuscript, EFT and KS revised the manuscript after peer-review.